\documentclass[prd,aps,preprint,tightenlines,showpacs,nofootinbib,superscriptaddress]{revtex4-1}
\usepackage{mathrsfs}
\usepackage{amsfonts}
\usepackage{amsmath}
\usepackage{amssymb}
\usepackage{array}
\usepackage{verbatim}
\usepackage{bm}
\usepackage{epsfig}
\usepackage{graphicx,color}
\usepackage{relsize}
\usepackage{lineno}
\usepackage{float}
\usepackage{multirow}
\RequirePackage{xspace}

\begin{document}

\begin{flushright}
MS-TP-23-26
\end{flushright}

\title{\boldmath Production of fully - heavy tetraquark states through the double parton scattering mechanism in $pp$ and $pA$ collisions}

\author{L. M. { Abreu}}
\email{luciano.abreu@ufba.br}
\affiliation{{ Instituto de F\'isica, Universidade Federal da Bahia, Campus Universit\'ario de Ondina, 40170-115, Bahia, Brazil}} 
\affiliation{{Instituto de F\'{\i}sica, Universidade de S\~{a}o Paulo, Rua do Mat\~ao, 1371, CEP 05508-090,  S\~{a}o Paulo, SP, Brazil}}

\author{F. { Carvalho}}
\email{babi.usp@gmail.com}
\affiliation{Departamento de Ci\^encias Exatas e
  da Terra, Universidade Federal de S\~ao Paulo,\\  
  Campus Diadema, Rua Prof. Artur Riedel, 275, Jd. Eldorado,
  09972-270, Diadema, SP, Brazil.}

\author{{J. V. C. { Oliveira} } }
\email{cerqueirs@gmail.com}
\affiliation{ Instituto de F\'isica, Universidade Federal da Bahia, Campus Universit\'ario de Ondina, 40170-115, Bahia, Brazil}

\author{V. P. { Gon\c{c}alves}}
\email{barros@ufpel.edu.br}
\affiliation{Institut f\"ur Theoretische Physik, Westf\"alische Wilhelms-Universit\"at M\"unster,
Wilhelm-Klemm-Stra\ss e 9, D-48149 M\"unster, Germany}
%\affiliation{Institute of Modern Physics, Chinese Academy of Sciences,
%  Lanzhou 730000, China}
\affiliation{Physics and Mathematics Institute, Federal University of Pelotas, \\
  Postal Code 354,  96010-900, Pelotas, RS, Brazil}

\begin{abstract}
The production of fully - heavy  tetraquark states in proton - proton ($pp$) and proton - nucleus ($pA$) collisions at the center - of - mass energies of the Large Hadron Collider (LHC) and at the Future Circular Collider (FCC) is investigated considering that these states are produced through the double parton scattering mechanism. We estimate the cross sections for the $T_{4c}$, $T_{4b}$ and $T_{2b2c}$ states and present predictions for $pp$, $pCa$ and $pPb$ collisions considering the rapidity ranges covered by central and forward detectors. We demonstrate that the cross sections for $pA$ collisions are enhanced in comparison to the $pp$ predictions scaled by the atomic number. Moreover, our results indicate that a search of these exotic states is, in principle, feasible in the future runs of the LHC and FCC. 
\end{abstract}

\maketitle

%\section{Introduction}

Over the last years, the LHCb \cite{Aaij:2020fnh}, ATLAS \cite{ATLAS:2023bft} and CMS \cite{CMS:2023owd} Collaborations have observed a sharp peak in the di - $J/\psi$ channel consistent with a narrow resonance at  $M = 6.9$ GeV, which is a viable candidate  for a fully - charm tetraquark state (for reviews see, e.g. Refs. \cite{Karliner:2017qhf,Olsen:2017bmm,Liu:2019zoy}).
Such data have motivated a series of studies that propose the existence of a large number of new exotic states,  composed only by charm and/or bottom quarks, denoted fully - heavy 
tetraquark states $T_{4Q}$ (See e.g. Refs. 
\cite{Debastiani:2017msn,Bedolla:2019zwg,Giron:2020wpx,Chen:2020xwe,Chao:2020dml,Lu:2020qmp,liu:2020eha,Lu:2020cns,Wang:2020ols,Becchi:2020uvq,
Becchi:2020mjz,Karliner:2020dta,Yang:2020rih,Weng:2020jao,Cao:2020gul,Yang:2020wkh,Wang:2020tpt,Faustov:2020qfm}).
Although the mass spectra and decay properties of these states are reasonably well understood, the production mechanism of fully - heavy tetraquark states is still a theme of intense debate (See e.g. Refs.   \cite{Karliner:2016zzc,Berezhnoy:2011xy,Carvalho:2015nqf,Esposito:2018cwh,Bai:2016int,Wang:2020gmd,Maciula:2020wri,Feng:2020riv,  Ma:2020kwb,Zhu:2020xni,Feng:2020qee,Goncalves:2021ytq,Biloshytskyi:2022dmo,Feng:2023agq}).
In particular,  Ref. \cite{Carvalho:2015nqf} has proposed that a fully - charm tetraquark state can be generated from the hadronization of   $c\bar{c}$ pairs, which are largely produced in double parton scatterings present in $pp$ collisions at LHC energies \cite{Luszczak:2011zp,Berezhnoy:2012xq,Cazaroto:2013fua}.
Such  an idea was 
 elaborated in more detail in Ref. \cite{Maciula:2020wri}, which confirmed that this mechanism is one of the more promising ways to probe the $T_{4c}$ state. One of the goals of this letter is the update of 
 Ref. \cite{Carvalho:2015nqf} in two aspects: (a) by assuming that the $T_{4c}$ state has a mass equal to 6.9 GeV instead of 5.4 GeV used in \cite{Carvalho:2015nqf}; and (b) by considering a more recent parametrization for the gluon distribution function. Another goal is to extend the model for the  fully - bottom tetraquark state, $T_{4b}$, and for the $c\bar{c} b\bar{b}$ state, denoted $T_{2b2c}$ hereafter, and present predictions for the associated cross sections derived considering $pp$ collisions at the LHC and FCC energies and assuming the typical rapidity ranges covered by central ($-2.5 \le Y \le +2.5$) and forward ($+2.0 \le Y \le +4.5$) detectors.
 Finally, our third and main goal, is to present for the first time the predictions for the $T_{4Q}$ production in proton - nucleus ($pA$) collisions via the double scattering mechanism. Our analysis is strongly motivated by the recent LHCb results \cite{LHCb:2020jse}, which observed the enhancement of the double scattering mechanism in $pA$ collisions predicted in Refs. \cite{treleanistrik,dps_pa,salvini,Cazaroto:2016nmu,Helenius:2019uge}. Assuming two distinct nuclei, we will estimate the energy dependence of the total cross sections and present predictions for future $pA$ collisions at the LHC and FCC. As we will demonstrate in what follows, our results indicate that if the fully - heavy tetraquark states are produced via the double parton scattering mechanism, the search of these states in proton - nucleus collisions is highly recommended.

\begin{figure}[t]
\includegraphics[scale=0.3]{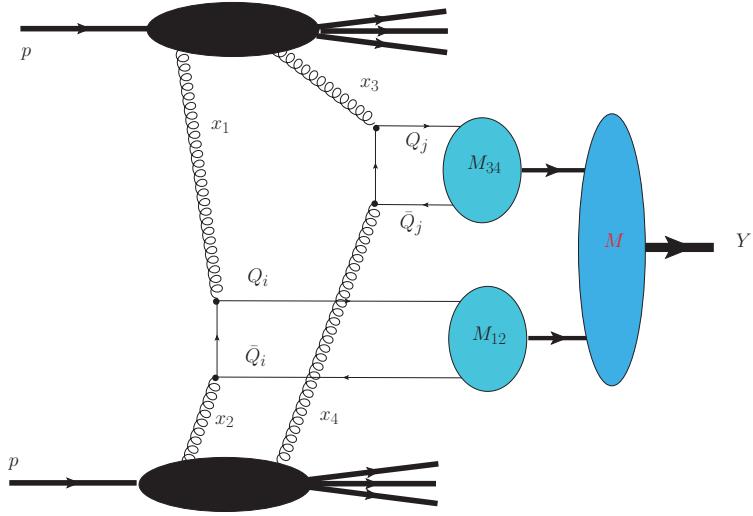}
\caption{Production of fully - heavy tetraquark states via the double parton scattering mechanism in $pp$ collisions.}
\label{Fig:diagram_pp}
\end{figure}

Initially, let us present a brief  review of the model proposed in  Ref. \cite{Carvalho:2015nqf} for the $T_{4Q}$ production in $pp$ collisions, which is represented in Fig. \ref{Fig:diagram_pp}. At the LHC energy,  
 the high density of partons in the hadron wave function implies the increasing of the probability that two or more hard partonic scatterings in a single hadron -- 
hadron collision can take place. Such theoretical expectation has been confirmed by distinct experimental collaborations at the LHC, considering different final states (See e.g. Ref. \cite{review}). In particular, the LHCb collaboration has observed a large number of events with four charm quarks ($c \bar{c} c \bar{c}$) in the same event \cite{LHCb:2012aiv}, which indicated a large contribution of   double parton scatterings (DPS). In Ref. \cite{Carvalho:2015nqf}, the authors have proposed that a final state composed by two heavy quark pairs, $Q_i \bar{Q}_i Q_j \bar{Q}_j$ can be generated by two independent gluon - gluon scatterings\footnote{The contribution of the $ q \bar{q} \rightarrow Q_i \bar{Q}_i$, where $q$ is a light quark, is subleading at the LHC energies.}, i.e. two times the reaction $ g g \rightarrow Q_i \bar{Q}_i$. These two pairs have invariant masses $M_{12}$ and  $M_{34}$, and are assumed to form a system with mass $M = M_{12} + M_{34}$ when the rapidities of the two $Q\bar{Q}$ pairs are equal, i.e.  $y_{12} = y_{34} = Y$.	The formation of the color neutral $T_{4Q}$ state is described by the Color Evaporation Model (CEM)\cite{cem,ramona}, which assumes that the color neutralization occurs through the emission of soft gluons and that the bound state is formed when the invariant mass of the  $Q_i \bar{Q}_iQ_j \bar{Q}_j$ system, $M$,  is of the order of  $M_{T_{4Q}}$.  
The  probability for the  $Q_i \bar{Q}_iQ_j \bar{Q}_j \rightarrow T_{4Q}$ transition is described in the CEM by a nonperturbative parameter $F_{{\cal{T}}}$. As detailed in Ref. \cite{Carvalho:2015nqf}, these assumptions imply that the production cross section can be expressed as follows
\begin{eqnarray}
\sigma_{T_{4Q}} (\sqrt{s}) =  &\,& \frac{F_{\cal{T}}}{\sigma_{eff,pp}}
\left[ \frac{1}{s} \int  d y_{12} \int dM_{12}^2 \,\,
g(\bar{x_1},\mu^2) \, g(\bar{x_2},\mu^2) \, \sigma _{g_1  g_2 \to Q_i\bar{Q}_i} \right] \nonumber \\
  & \times & \,\,
\left[ \frac{1}{s} \int  dy_{34} \int  dM_{34}^2 \,\,
g(\bar{x_3},\mu^2) \, g(\bar{x_4},\mu^2) \, \sigma _{g_3 g_4 \to Q_j\bar{Q}_j} \right] \nonumber \\
  & \times & \,\, \Theta (1- \bar{x_1} - \bar{x_3}) \,\,\Theta (1- \bar{x_2} - \bar{x_4}) \,\,
\Theta (M_{12}^2 - 4m_{Q_i}^2) \,\, \Theta (M_{34}^2 - 4m_{Q_j}^2) \nonumber \\
& \times & \,\, \delta(y_{34} - y_{12})
\label{sigtotym}
\end{eqnarray}
where $\sigma_{eff,pp} = 15$ mb is a constant extracted from data analysis,  $g(x,\mu^2)$ is the gluon distribution in the proton with a fractional momentum 
$x$  at the factorization scale $\mu^2$ and $\sigma _{g g \to Q_i\bar{Q}_i}$ is the 
 elementary cross section for the $g g\to Q_i\bar{Q}_i$ subprocess. The  step functions $\Theta (1- x_1 - x_3)$ and
$\Theta (1- x_2 - x_4)$    enforce momentum conservation in the projectile and in the target, while the step functions $\Theta (M_{12}^2 - 4m_{Q_i}^2)$ and $\Theta (M_{34}^2 - 4m_{Q_j}^2)$   guarantee 
that the invariant masses of the gluon pairs 12 and 34 are large enough to produce two heavy 
quark pairs. Moreover, the delta function implements the  condition that the two heavy quark pairs are in the same rapidity. The Bjorken variables of the four gluons in the initial state are given by
\begin{equation}
\bar{x_1} = \frac{M_{12}}{\sqrt{s}} \, e^{y_{12}}   \,\, \hspace{0.3cm} , \hspace{0.3cm}
\bar{x_2} = \frac{M_{12}}{\sqrt{s}} \, e^{- y_{12}} \,\, \hspace{0.3cm} , \hspace{0.3cm}
\bar{x_3} = \frac{M_{34}}{\sqrt{s}} \, e^{y_{34}}   \,\, \hspace{0.3cm} , \hspace{0.3cm}
\bar{x_4} = \frac{M_{34}}{\sqrt{s}} \, e^{- y_{34}} \,\,.
\label{redefex}
\end{equation}
In our calculations we will assume $m_c = 1.5$ GeV, $m_b = 4.5$ GeV  and that the hard scale $\mu$ is equal to the invariant mass of the $Q\bar{Q}$ system. Moreover, we will consider the CT14 parametrization \cite{Gao:2013xoa} for the gluon distribution function. The last ingredient needed to estimate the $T_{4Q}$ production in $pp$ collisions is the value of $F_{\cal{T}}$ for the $T_{4c}$, $T_{4b}$ and $T_{2b2c}$ states. As in Ref. \cite{Carvalho:2015nqf}, its value for the $T_{4c}$ production will be estimated in terms of the cross section for the $X(3872)$ production measured by the CMS collaboration \cite{cms} in $pp$ collisions at $\sqrt{s} = 7.0$ TeV.  The value of $F_{\cal{T}}$ is determined by imposing that $\sigma_{T_{4c}} = 0.12 \sigma_X$, which implies $F_{\cal{T}} = 0.00119$. In what follows, we will assume this same value in the calculation of the $T_{4b}$ and $T_{2b2c}$ production cross sections. Such assumption is motivated by the results obtained e.g. in Refs. \cite{GayDucati:2001bf,BrennerMariotto:2001sv}, which have  indicated that the value of this nonperturbative factor is similar for different quarkonium states.

\begin{figure}[t]
\includegraphics[scale=0.23]{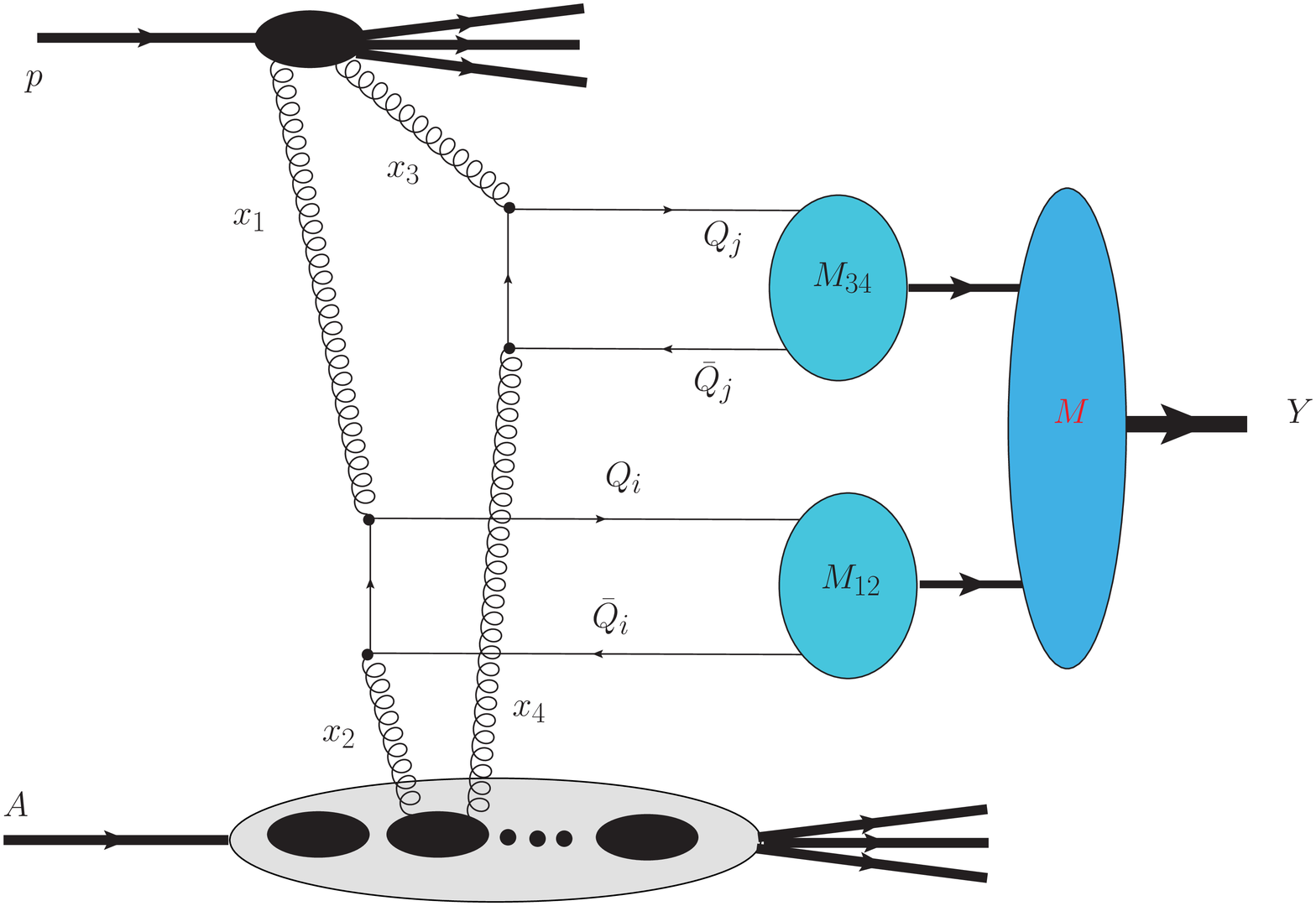}
\includegraphics[scale=0.23]{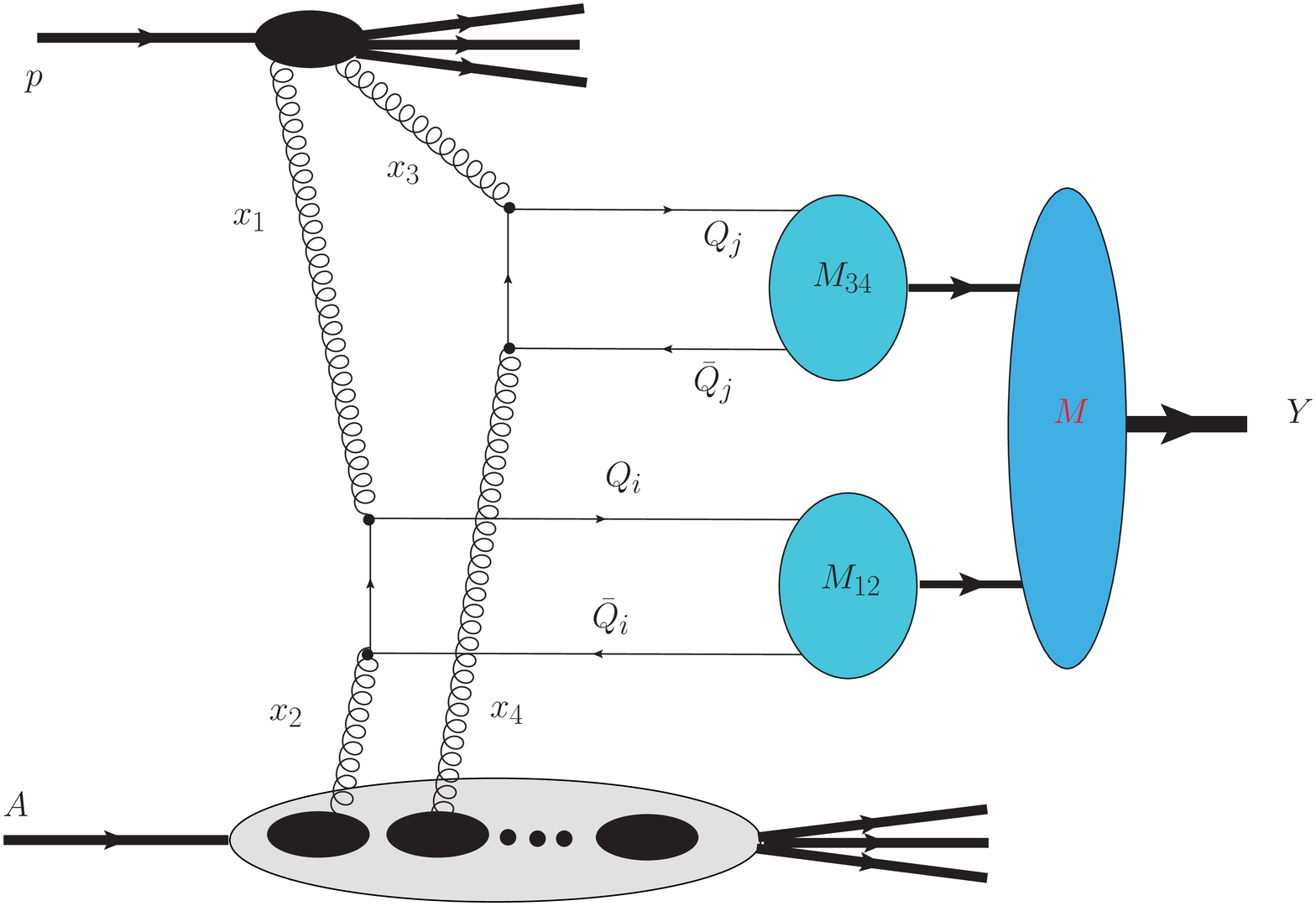}
\caption{Representation of the two diagrams that contribute for the production of fully - heavy tetraquark states via the double parton scattering mechanism in $pA$ collisions.}
\label{Fig:diagram_pA}
\end{figure}

In order to generalize the model for  proton - nucleus collisions, we should  take into account that the parton flux associated to the nucleus is enhanced by a factor $\propto A$ and that the two gluons associated to the proton can interact  with two gluons coming from the same nucleon from the nucleus  or with two gluons coming from different nucleons from the nucleus. Both possibilities are 
represented in the left and right panels of the Fig  \ref{Fig:diagram_pA}. In what follows, we will denote  the cross sections associated to these two contributions by   $\sigma_{pA}^{DPS,1}$ and  $\sigma_{pA}^{DPS,2}$, respectively.  
Following Refs. \cite{treleanistrik,dps_pa}, one has that $\sigma_{pA}^{DPS,1}=A\cdot \sigma_{pp}^{DPS}$. Moreover, cross section associated to the second contribution will be given by  
$\sigma_{pA}^{DPS,2}=\sigma_{pp}^{DPS}\cdot \sigma_{eff,pp}\cdot F_{pA}$, with 
$F_{pA}=[(A-1)/A]\,\int T^2_{pA}({\bf r})d^2r$, where  ${\bf r}$ is the 
impact parameter between the colliding proton and nucleus and $T_{pA}$ is the  nuclear 
thickness function. Assuming  that the nucleus has a spherical form (with uniform nucleon density) of radius $R_A = r_0A^{1/3}$, and $r_0=1.25$ fm, the 
integral of the nuclear thickness factor becomes $F_{pA} = {9A(A-1)}/{(8\pi R_A^2)}$.
As a consequence, the formalism proposed in Ref. \cite{treleanistrik} implies that the cross section for the production of fully - heavy tetraquark states via the double parton scattering mechanism in $pA$ collisions can be expressed as follows 
\begin{eqnarray}
\sigma^{DPS}_{pA \to T_{4Q}}  & = &  \sigma_{pA \to T_{4Q}}^{DPS,1} + \sigma_{pA\to T_{4Q}}^{DPS,2} \nonumber \\
& = & A \sigma^{DPS}_{pp \to \to T_{4Q}} \left[1 + \frac{1}{A} \sigma_{eff,\,pp}F_{pA}\right]  \,\,,
\label{sig_dps_pa1}
\end{eqnarray}
where $\sigma^{DPS}_{pp \to \to T_{4Q}}$ is given by Eq. (\ref{sigtotym}).
In our calculations, we will consider two distinct nuclei ($A = 40$ and 208).

\begin{figure}[th!]
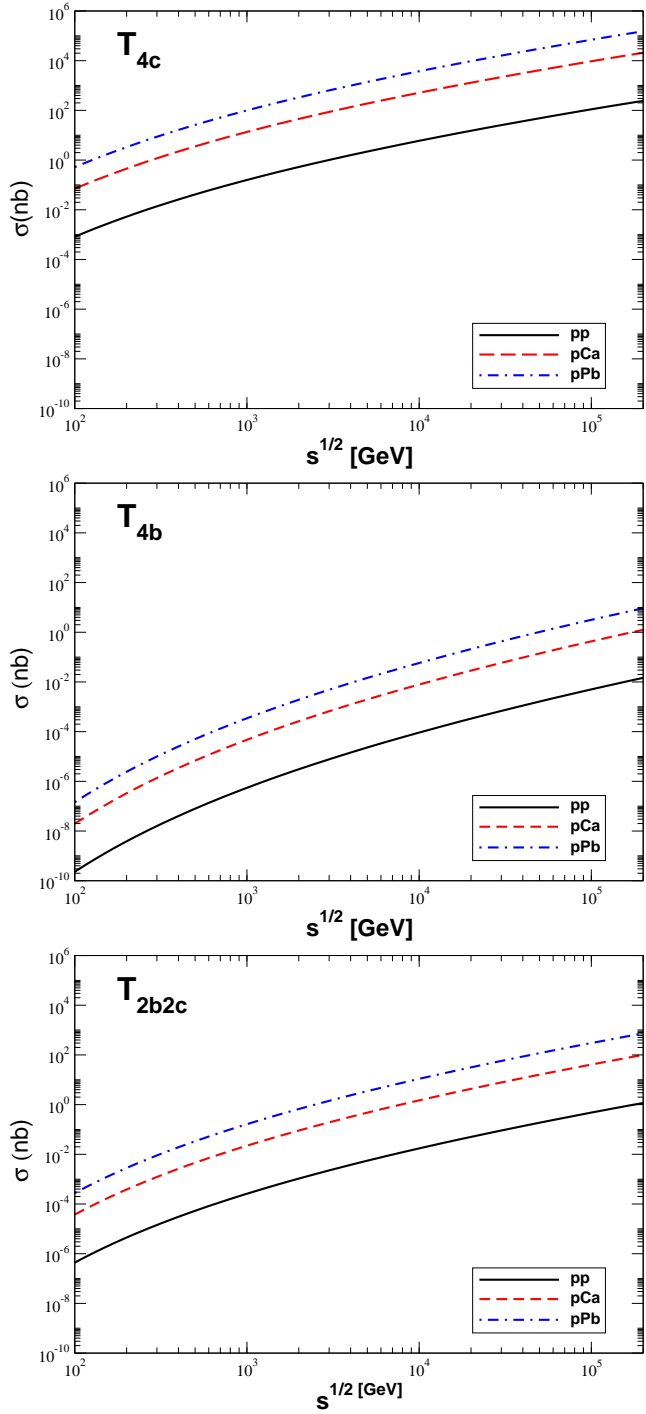

\includegraphics[scale=0.35]{T4c.eps}
\includegraphics[scale=0.35]{T4b.eps} 
\includegraphics[scale=0.35]{T2bc.eps}  
%\\
%(c) & \, & (d) 
%\end{tabular}
%\includegraphics[scale=0.42]{minv_ea_central_Pbp.pdf}
%\includegraphics[scale=0.42]{minv_ea_frontal_Pbp.pdf}
\caption{Energy dependence of the cross sections for the production of fully - heavy tetraquark states via the double parton scattering mechanism in $pp$, $pCa$ and $pPb$ collisions.}
\label{Fig:energy}
\end{figure} 

In Fig. \ref{Fig:energy} we present our predictions for the energy dependence of the cross sections for the production of fully - heavy tetraquark states via the double parton scattering mechanism in $pp$, $pCa$ and $pPb$ collisions. A strong increasing with the energy is predicted, which is directly associated with the fact that in our model for the $T_{4Q}$ production, the cross section is proportional to $g(x,\mu^2)^4$. In contrast, in models based on the single parton scattering mechanism, one has $\sigma^{SPS}_{T_{4Q}} \propto g(x,\mu^2)^2$. Therefore, we can expect that for large energies one will have $\sigma^{DPS}_{T_{4Q}} > \sigma^{SPS}_{T_{4Q}}$. The results derived in \cite{Luszczak:2011zp,Cazaroto:2013fua} for the four charm production indicate that DPS charm 
production is already comparable to SPS production at LHC energies. As a consequence, our predictions for the $T_{4Q}$ production can be considered a lower bound for the number of events in hadronic collisions at the LHC and FCC.

For the $T_{4c}$ production  (upper panel), one has that our predictions for $pp$ collisions are similar to those derived in Ref. \cite{Carvalho:2015nqf}, which is directly associated to the fact that in both studies the normalization was fixed using the cross section for the $X(3872)$ production. On the other hand, our prediction for $pCa$ collisions is a factor $\approx 85$ larger than for the $pp$ case, i.e. it is not a simple $A$ scaling of the $pp$ prediction. Such  result is expected due to contribution of $\sigma_{pA}^{DPS,2}$ (right panel in Fig. \ref{Fig:diagram_pA}). Similarly, the predictions for $pPb$ collisions are enhanced by a factor $\approx 630$, i.e. $\approx 3\,A$, in agreement with the results derived in Ref. \cite{Cazaroto:2016nmu}, where the double heavy quark pair production in $pA$ collisions has been estimated for the first time. 

Our predictions for the $T_{4b}$ and $T_{2b2c}$ production are presented in the middle and lower panels of Fig. \ref{Fig:diagram_pA}, respectively. One has that the energy behaviour is similar to that predicted for the $T_{4c}$ case and that the $pCa$ and $pPb$ cross sections are enhanced by a similar factor in comparison to the $pp$ predictions.
The main difference is in the magnitude of the cross sections. In Tables \ref{Tab:lhc} and \ref{Tab:fcc} we present our predictions for the $T_{4Q}$ production cross sections at the LHC and FCC energies, respectively, considering the typical rapidity ranges covered by a central ($-2.5 \le Y \le +2.5$) and forward ($+2.0 \le Y \le +4.5$) detectors.  We predict cross sections of the order of few nb for the $T_{4c}$ production in $pp$ collisions at the LHC. In contrast, for the $T_{4b}$ ($T_{2b2c}$) case, the values predicted are smaller by four (two) orders of magnitude. At the FCC, we predict an increasing in the cross sections of one order of magnitude, which is directly associated to the fact that $\sigma^{DPS}_{T_{4Q}} \propto xg^4$.  
For $pA$ collisions, one has the enhancement  discussed in the previous paragraph.
These results indicate that the cross sections for the $T_{4Q}$ production in $pp$ and $pA$ collisions are large and that a future experimental analysis is, in principle, feasible.

 \begin{table}[t]
\begin{center}
\begin{tabular}{ |c|c|c|c|}

\hline
 & {\bf pp ($\sqrt{s} = 14.0$ TeV)} & {\bf pCa ($\sqrt{s} = 8.1$ TeV)} & {\bf pPb ($\sqrt{s} = 8.1$ TeV)} \\
\hline 
 & \begin{tabular}{c|c}
    {\bf Central} & {\bf Forward}      
\end{tabular}  
 &\begin{tabular}{c|c}
    {\bf Central} & {\bf Forward}      
\end{tabular} 
 & \begin{tabular}{c|c}
    {\bf Central} & {\bf Forward}      
\end{tabular} \\

\hline

{\bf $T_{4c}$} 
&   \begin{tabular}{c|c}
      6.05 & 1.86 
  \end{tabular}  
       & 
 \begin{tabular}{c|c}
       520.22 & 159.74  
  \end{tabular} 
       & 
  \begin{tabular}{c|c}
       3816.44 & 1171.86  
  \end{tabular} \\

\hline

{\bf $T_{4b }$} & 
\begin{tabular}{c|c}
       0.00033 & 0.000078 
  \end{tabular}  
       & 
 \begin{tabular}{c|c}
       0.028 & 0.0067   
  \end{tabular} 
       & 
  \begin{tabular}{c|c}
       0.21 & 0.049   
  \end{tabular} \\
\hline

{\bf $T_{2b2c}$} & 
\begin{tabular}{c|c}
       0.021 & 0.0055  
  \end{tabular}  
      & 
 \begin{tabular}{c|c}
       1.80 & 0.47
  \end{tabular} 
       & 
  \begin{tabular}{c|c}
       13.23 & 3.46
  \end{tabular} \\

\hline
%\caption{pp 14TeV e pA 8.1TeV}
\end{tabular}
\caption{Cross sections in nb for the $T_{4Q}$ production in $pp$ and $pA$ collisions at the LHC calculated considering the rapidity ranges covered by a typical central ($-2.5 \le Y \le +2.5$) and forward ($+2.0 \le Y \le +4.5$) detectors.}
\label{Tab:lhc}
\end{center}
\end{table}

\begin{table}[t]
\begin{center}
\begin{tabular}{ |c|c|c|c|}

\hline
 & {\bf pp ($\sqrt{s} = 100.0$ TeV)} & {\bf pCa ($\sqrt{s} = 63.0$ TeV)} & {\bf pPb ($\sqrt{s} = 63.0$ TeV)} \\
\hline 
 & \begin{tabular}{c|c}
    {\bf Central} & {\bf Forward}      
\end{tabular}  
 &\begin{tabular}{c|c}
    {\bf Central} & {\bf Forward}      
\end{tabular} 
 & \begin{tabular}{c|c}
    {\bf Central} & {\bf Forward}      
\end{tabular} \\

\hline

{\bf $T_{4c}$} & 
  \begin{tabular}{c|c}
       57.27 & 21.74  
  \end{tabular}  
       & 
 \begin{tabular}{c|c}
       4918.35 & 1867.03  
  \end{tabular} 
       & 
  \begin{tabular}{c|c}
       36081.93 & 13696.89   
  \end{tabular} \\

\hline

{\bf $T_{4b}$} & 
\begin{tabular}{c|c}
       0.0081 & 0.0026
  \end{tabular}  
       & 
 \begin{tabular}{c|c}
       0.70 & 0.22 
  \end{tabular} 
       & 
  \begin{tabular}{c|c}
       5.11 & 1.63
  \end{tabular} \\
\hline

{\bf $T_{2b2c}$} & 
\begin{tabular}{c|c}
       0.27 & 0.095 
  \end{tabular}  
       & 
 \begin{tabular}{c|c}
       23.18 & 8.16  
  \end{tabular} 
       & 
  \begin{tabular}{c|c}
       170.11 & 59.85  
  \end{tabular} \\

\hline
%\caption{pp 100 Tev e pA 63 TeV
\end{tabular}
\caption{Cross sections in nb for the $T_{4Q}$ production in $pp$ and $pA$ collisions at the FCC calculated considering the rapidity ranges covered by a typical central ($-2.5 \le Y \le +2.5$) and forward ($+2.0 \le Y \le +4.5$) detectors.}
\label{Tab:fcc}
\end{center}
\end{table}

As a summary, in this letter we have investigated the production of fully - heavy tetraquark states through the double parton scattering (DPS) mechanism in $pp$ and $pA$ collisions at the LHC and FCC. Such contribution is expected to be, at least, of the same order of the contribution associated to the $T_{4Q}$ production via the single parton scattering, which implies that our predictions can be considered a lower bound for the magnitude of the total cross sections. In our analysis, one has updated the predictions for the $T_{4c}$ production in $pp$ collisions presented in Ref.  \cite{Carvalho:2015nqf} and extended the model for the production of the  $T_{4b}$ and $T_{2b2c}$ states. Our results indicated that the cross sections for these states are, respectively, four and two orders of magnitude smaller than the $T_{4c}$ predictions. However, considering the large luminosity expected in the future runs of the LHC, one predict a large number of events, which makes possible to search for these states in the forthcoming years.
In addition, in this letter one has presented, for the first time, the predictions for the $T_{4Q}$ production in $pCa$ and $pPb$ collisions. We have demonstrated that the DPS cross sections are enhanced by a factor larger than the expected $A$ scaling predicted by a model based on the simple superposition of proton - nucleon collisions. In particular, we predict that the $pCa$ ($pPb$) cross section will be enhanced by a factor $\approx 2\,A \,(3\,A)$ in comparison to the $pp$ result for the same center - of - mass energy. Such result indicates that  a future experimental analysis of the $T_{4Q}$ production in $pA$ collisions can be useful to probe the existence of these states, as well to improve our understanding of the double parton scattering mechanism.

\section*{Acknowledgments}
This work was partially supported by  CNPq, CAPES, FAPERGS, FAPESB and  INCT-FNA (process number 
464898/2014-5).

\end{document}